\documentclass[aps,pra,groupedaddress,amssymb,amsmath,showpacs,twocolumn]{revtex4}
\usepackage{graphicx}

\begin{document}

\title{High temperature phase transition in the coupled atom-light system in the presence of  optical collisions}
\author{A. P. Alodjants}
\email[]{alodjants@vlsu.ru}
\author{I. Yu. Chestnov}
\author{S. M. Arakelian}
\affiliation{Department of Physics and Applied Mathematics, Vladimir State University, Gorky str. 87, 600000, Vladimir, Russia}
\pacs{{42.50.Nn}, {05.30.Jp}}

\begin{abstract}
The problem of  photonic phase transition for the system of a two-level atomic ensemble interacting with a quantized single-mode electromagnetic field in the presence of optical collisions (OC) is  considered.  We have shown that  for  large and negative atom-field detuning a photonic field exhibits high temperature second order phase transition  to superradiant state under thermalization  condition for coupled atom-light states. Such a transition can be connected with superfluid (coherent) properties of photon-like low branch (LB) polaritons. We discuss the application of metallic cylindrical waveguide for observing predicted effects. 
\end{abstract}
\maketitle

\section{INTRODUCTION}\label{sec_I}

Nowadays the investigation of phase transitions in atomic gases represents a huge area of experimental and theoretical research where condensed matter and statistical physics are closely connected to the application problems of quantum and atom optics, e.g., in the field of quantum information science -- see e.g. \cite{1}. Although Bose-Einstein condensation (BEC) of the atoms has been observed in many labs, the requirement to use extremely low (up to microKelvins) temperatures strictly limits the utilization of such an effect for practical purposes. It provides an important reason for studying relatively high temperature phase transitions. Usually, such transitions take place in coupled matter-field systems -- polaritons introduced many years ago for describing the interaction of quantized field with quantum excitations in the medium, see e.g.~\cite{2}. At present the evidence of phase transitions with LB polaritons and their superfluid properties have been observed  in solid state physics with semiconductor microstructures -- see e.g. \cite{3,4,5}. In such systems  polaritons can be treated as 2D gas of bosonic particles  -- so-called exciton-polaritons appearing  in the sample of quantum wells inserted in a semiconductor (CdTe/CdMgTe or GaAs) microcavity and  having effective mass which is many orders smaller than a free mass of  electrons.  However, high (room)  temperature  phase transition for current \textit{narrow-band} semiconductors  seems to be hard to reach due to exciton ionization. At the same time the time of thermalization for polaritons in solid state structures mentioned above is short enough and is in the picoseconds regime \cite{3}. In this sense polaritons in atomic physics could be preferable for observing high-temperature phase transitions. Such polaritons potentially possess a longer coherence time. In particular,  atomic polaritons represent superposition of photon and polarization of two (or multi)-level atoms  and can be  observed in various problems of  atom-field interaction where long lived coherence  of  quantized optical field strongly coupled with macroscopic atomic ensemble plays an essential role -- cf.~\cite{6,7}. 

The primary step for the experimental observation of  phase transitions and BEC for atomic polaritons is connected with  achieving thermal equilibrium in the coupled atom-light system.  The so-called optical collisions (OC) have been proposed recently  for this purpose. The process of OC represents nonresonant interaction of a quantized light field with an atom in the presence of buffer gas particle -- see e.g.~\cite{8}. Although the main features of OCs have been investigated both in theory and in experiment for a long time (see e.g.~\cite{9,10,11}), the thermodynamic properties of a coupled atom-light system have been studied quite recently \cite{12,13,14,32,15}. In particular, in \cite{14} it has been demonstrated for the first time that OCs lead to the thermalization of coupled (dressed) atom-light states under the interaction of the optical field with rubidium atoms in an ultrahigh-pressure buffer gas cell at high (530 K) temperature. Latterly  in \cite{15}  we have proposed a theoretical approach  of a dressed-state thermalization that accounts  the evolution of pseudo-spin Bloch vector components and characterizes the essential role of the spontaneous emission rate in the thermalization process.  Although we have used a two-level model for describing OC process  with buffer gas particles (see e.g. \cite{9,10,11}) the  dependences theoretically obtained  in \cite{15} are qualitatively in good agreement with  experimentally observed results  for rubidium atoms.  The predicted time of thermalization was  in a nanosecond domain at full optical power 300~mW and negative atom-light detuning $\delta / 2\pi = -11$~THz. The main features of such a thermalization are connected with the observed asymmetry of a saturated lineshape that depends on the temperature of atomic gas and  detuning  $\delta =\omega _{L} -\omega _{at} $; $\omega _{L} $ and  $\omega _{at} $ are frequencies of optical field and atomic transition respectively.  

In a number of papers  \cite{12,13}  authors discussed  thermodynamic properties of a two-level sodium atomic system being under OCs with buffer gas particles in another limit of  positive atom-light detuning, i.e. for $\delta >0$. In particular, there has been shown a strong laser generation in the atomic ensemble  interacting with cw-laser field in the domain of a far blue wing of spectral line under the frequent collision with buffer gas particles. In this sense both problems, i.e. lasing that occurs in  atomic system in the presence of OC and phase transition for coupled atom-light states, evoke great interest and should be clarified. 

Physically the problems under discussion  are very close to another intriguing and long stated problem of phase transition for a light field (or BEC of photons) - see, e.g.,~\cite{16,17,18,19}. In fact, the comparison of lasing and condensation phenomena for photonic, atomic and solid state systems as two possible ways for achieving macroscopic coherence has long been discussed -- see e.g.~\cite{20,21,22}. The first attempts to consider some analogy between lasing under the threshold region and second-order phase transition in ferromagnets have been made in \cite{20}. Later a simple Dicke model that describes an ensemble of two-level atoms interacting with quantized electromagnetic field in standard laser quantum theories has been proposed for observing phase transition of photons \cite{16}. In quantum optics phase transition in such a system has been interpreted as a transition to some superradiant (coherent) photonic state with zero chemical potential \cite{17}. At the same time it is pointed out in \cite{18} that in this case the establishment of spontaneous static field (a field with zero frequency) in the medium takes place. In fact such a phase transition creates some ferroelectric state in the medium. 

From our point of view there exist important circumstances that must be taken into account while considering phase transitions mentioned above. In many cases including usual lasers, we deal in practice with \textit{non-equilibrium }states of the system, cf.~\cite{21,24}. On the other hand,  Bose-Einstein condensation phenomenon and phase transition under discussion have some meaning only in the limit of thermodynamically  equilibrium state of a coupled atom-light system for which chemical potential is nonzero, cf.~\cite{25}.  In this sense the applicability of a convenient Dicke model for the phase transition problem should be justified in each physical case.

Noticing, that in previous experiments \cite{14,15} the region of high and uniform laser intensity was about 70 $\mu m$. Physically it means that the medium is thin and the lifetime of photon-like polaritons discussed in this paper seems to be short in comparison with the thermalization time of a coupled atom-light system. However, the implementation of special metallic waveguides of various configurations with the length up to a few mm's in the OCs experiment performed recently (see \cite{32}) allows to increase the time of atom-field interaction by many orders due to photon trapping and confinement. In fact, such waveguides pave the way to the investigation of the high temperature phase transition problem with polaritons in atomic optics.

In the present paper we continue our theoretical investigation of thermodynamic  and critical properties of coupled atom-light states appearing due to the interaction of  two-level atoms  with a single-mode optical field in the presence of  buffer gas particles. In   Sec.\ref{sec_II}  we   examine critical properties of photonic field under the  thermalization of coupled atom-light states. In Sec.\ref{sec_III} we study the problem of a second order phase transition  in the system under discussion.  In particular, we suggest a simple  model of polaritons describing  atom-field interaction at equilibrium. Relying on a thermodynamical approach we consider a mean-field Bardeen-Cooper-Schrieffer (BCS)-like gap  equation for order parameter -- normalized amplitude of photonic field.  In Appendix we discuss the properties of a cylindrical  waveguide for realizing appropriate strong atom-field coupling in a single-mode regime.   In conclusion, we summarize the  results obtained.

\section{THERMODYNAMIC APPROACH FOR PHOTONIC FIELD UNDER THE OC PROCESS} \label{sec_II}

\begin{figure}
\includegraphics[scale=0.4]{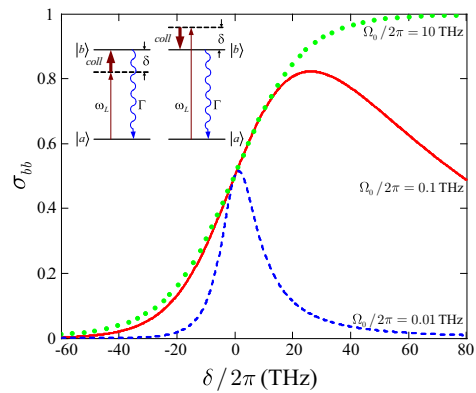}
\caption{\label{fig1}  (Color online) Population of the upper state $\sigma _{bb} $ as a function of  atom-field detuning $\delta /2\pi $ for 500-bar argon buffer gas at different values of the resonant Rabi frequency $\Omega _{0}^{} /2\pi $. The parameters are: $\gamma/2\pi = 3.6$ THz, ${\Gamma} \simeq 2\pi\cdot6$~MHz. The inset includes the scheme of   collisionally aided absorption in a two-level atom  for  $\delta < 0$ (left panel) and for  $\delta > 0$ (right panel).}
\end{figure}

An optical collision (OC) is an elementary process of a collision between an isolated  atom of sort A and  a foreign (buffer) gas particle of sort B  resulting in the emission or absorption of non-resonant photons -- see e.g. \cite{8}. The inset of Fig.\ref{fig1} includes a two-level model for characterizing OC process under consideration. Actually, in recent experiments \cite{14,15,32} Rabi splitting frequency, atom-field detuning and collisional broadening are at least three and more orders larger than splitting frequency between hyperfine levels for 5S--5P transition in rubidium \textit{D}-lines, cf.~\cite{26}. As it is demonstrated in \cite{14,15} usual pressure-broadened rubidium $D_{1}$ and $D_{2}$ lines are visible at a moderate optical power $P = 25$ mW independently. In this limit collisional broadening can be treated for $D_{1}$ and $D_{2}$ lines of rubidium atoms separately. Contrary,  at full optical power 300 mW for which the  thermalization of  coupled atom-light states occurs the essential  asymmetry  of $D_{1}$ and $D_{2}$ spectral lines and a common shape of pressure broadening curves  are clearly seen in the experiment, cf. \cite{14,15,32}. Hence, the assumption of a two-level approach for OC problem neglecting hyperfine structure of rubidium \textit{D}-lines seems to be justified, cf.~\cite{13}.    

We suppose that  an optical field with frequency $\omega _{L} $ interacts non-resonantly with two-level atoms having frequency spacing $\omega _{at} $ in the presence of buffer gas particles. Physically OC means that  excitation of level ${\left| b \right\rangle} $ is impossible due to atomic collisions only. A theoretical description of coupled atom-light state thermalization  can be given by the density matrix approach, see e.g.~\cite{15}. In general, the interaction of two-level atoms with a quantized optical field in the presence of collisions with buffer gas particles is  characterized by the following physical parameters. First, there is    Rabi frequency  $\tilde{\Omega }_{R} =\Omega _{R} -{\delta \eta \mathord{\left/ {\vphantom {\delta \eta  \Omega _{R} }} \right. \kern-\nulldelimiterspace} \Omega _{R} } $ that describes the splitting of energy levels in the presence of collisions with buffer gas particles, $\Omega _{R} =\sqrt{\Omega _{0}^{2} +\delta ^{2} } $ is a definition of the same quantity without collisions when $\eta =0$;  $\Omega _{0} =2g\sqrt{N_{ph} } $ is resonant Rabi splitting taken at detuning $\delta =0$;  $g=\left(\frac{\left|d_{ab} \right|^{2} \omega _{L} }{2\hbar \varepsilon _{0} V} \right)^{1/2}$ is the atom-field interaction constant, which we assume to be identical for all atoms;  $d_{ab} $ is the atomic dipole matrix element; $V$ is the interaction volume; $N_{ph} $ is a total average number of photons involved in the OC process.  

The collisions with buffer gas particles are  determined by two parameters  $\eta $ and $\gamma $  which can be expressed phenomenologically  via the average phase shift  that is accumulated during the collision, cf.~\cite{8}. Physically parameters $\eta$ and $\gamma$ can be also connected with molecular potentials of a compound A$+$B system.  Parameter $\eta $ depends on the difference $\Delta U$ of diagonal matrix elements for interacting atoms A and B due to their collisions (scattering). According to the quantum mechanical approach to the OC problem, $\Delta U$ can be represented as $\Delta U=C_{n} R^{-n} $ for power law of atomic interaction; $C_{n} $ is a constant that depends on a character of  such interaction;  $R$ is a distance between the  atom and perturber (buffer gas particle) for a given time; for more details see~\cite{11}.  

Second, OC process is  described by collisional rate (collisional broadening)  $\gamma $ that plays an important role in thermalization process. In a general case parameter $\gamma $ is characterized by the density of buffer gas particles,  molecular potentials  for a compound system  and depends on the value of atom-field detuning  $\delta $ -- see~\cite{11}. In the experiment described in \cite{14}   parameters $\gamma $ and $\eta $ are of the order of  terahertz.    Skipping some  details of our calculations presented in \cite{15},  we are starting here with  important results related to thermalization of coupled (dressed) atom-field states.  

Figure \ref{fig1} demonstrates a calculated population of the upper state $\sigma _{bb} $ \textit{ }that is proportional to the total intensity of spectral components as a function of atom-field detuning  $\delta $\textit{ }taken for the interaction of a quantized field with  two-level rubidium atoms characterized by mean resonance frequency of transition 382 THz corresponding to a weighted mean of rubidium \textit{D}-lines, cf.~\cite{26}.  In the  paper we focus on the so-called  perturbative limit when atom-field detuning $\delta $ is large  enough,  i.e. when inequalities $\left|\delta \right|>\gamma ,\ \eta >\Omega _{0} $ are held.  In this limit one can neglect phase shift introduced by OC, assuming that $\tilde{\Omega }_{R} \approx \Omega _{R} $. At the same time the scaled atomic population of the upper level $\sigma _{bb} $ in this case yields 
\begin{equation} \label{eq1_} 
\sigma _{bb} \simeq \frac{1}{2\left(1+\frac{\Gamma \delta ^{2} }{\gamma \Omega _{0}^{2} } \right)} \left[1+\frac{\delta }{\left|\delta \right|} \tanh \left(\frac{\hbar \left|\delta \right|}{2k_{B} T} \right)\right],                                    
\end{equation} 
where $\Gamma $ is a spontaneous emission rate.

First, we examine the role of atomic collisions in the thermalization process assuming that ${\Gamma \delta ^{2} \mathord{\left/ {\vphantom {\Gamma \delta ^{2}  \gamma \Omega _{0}^{2} }} \right. \kern-\nulldelimiterspace} \gamma \Omega _{0}^{2} } \gg 1$; a blue dashed curve in Fig.\ref{fig1} corresponding to this limit.   For negative detuning $\delta <0$ from  \eqref{eq1_} we get  
\begin{equation} \label{eq2_} 
\sigma _{bb} \simeq \frac{\gamma \Omega _{0}^{2} }{\Gamma \delta ^{2} } e^{-{\hbar \left|\delta \right|\mathord{\left/ {\vphantom {\hbar \left|\delta \right| k_{B} T}} \right. \kern-\nulldelimiterspace} k_{B} T} } ,                                                                  
\end{equation} 
 where we also suppose that  inequality  
\begin{equation} \label{eq3_} 
\hbar \left|\delta \right|\gg k_{B} T 
\end{equation} 
is satisfied.  

The relation \eqref{eq3_} can be recognized as a ``low temperature'' limit in the framework of a convenient approach to existing theories for phase transitions and BEC problem, cf.~\cite{27}. However, in the case of OCs condition \eqref{eq3_} can be achieved for high enough temperatures and for very large atom-field detuning $\delta$. 

The result obtained from \eqref{eq2_} is in good agreement with theoretical predictions for the spectral line obtained for a far red wing (so-called adiabatic wing) -- see e.g.~\cite{11}. In particular, the curve of spectral intensity  decays exponentially in this case. 

Now let us switch over to a positive-valued large atom-field detuning, that is, to $\delta > 0$.    In this case  we get from \eqref{eq1_}
\begin{equation} \label{eq4_} 
\sigma _{bb} \approx \frac{\gamma \Omega _{0}^{2} }{\Gamma \delta ^{2} } .                                                                    
\end{equation} 

The  Eq.\eqref{eq4_} reproduces  a physical  result for a far blue wing (so-called,  static wing of spectral line) well known from the theory of OCs, cf.~\cite{11}. The result is that the population of the upper level is inversely proportional to square of atom-light  detuning.

However, according to our theoretical approach important features of  population of the upper state  are connected with the dependence of $\sigma_{bb}$ on temperature $T$ of atomic gas.  

The thermalization of coupled atom-light (dressed) states occurs if condition
\begin{equation} \label{eq5_} 
\Gamma /\gamma \ll \Omega ^{2}_{0} / \delta^{2} \ll 1
\end{equation} 
is fulfilled. In particular,  for large $\Omega _{0} $ the $\sigma _{bb} $  approaches Dirac-Fermi distribution  function 
\begin{equation} \label{eq6_} 
\sigma _{bb} \approx \frac{1}{1+e^{-{\hbar \delta \mathord{\left/ {\vphantom {\hbar \delta  k_{B} T}} \right. \kern-\nulldelimiterspace} k_{B} T} } }.                                                    
\end{equation} 

In \cite{14,15} the best result concerning  coupled atom-light state thermalization  has been experimentally observed for maximally accessible atom-field detuning $\delta/2\pi=-11$~THz and resonant Rabi frequency ${\Omega_{0}/2\pi=0.1}$~THz (indicated by solid curve in Fig.\ref{fig1}). The obtained time of thermalization was $10$ times shorter than the natural lifetime for rubidium \textit{D}-lines.

For further processing it is important to determine a total atom-field excitation  (polariton) density  $\rho$  as 
\begin{equation} \label{eq7_} 
\rho =\lambda _{}^{2} +\sigma _{bb},
\end{equation} 
that we consider to be constant at thermal equilibrium. In \eqref{eq7_} $\lambda^2 = \left\langle f^{\dag} f\right\rangle /N $ is  a normalized average photon number, $f$ ($f^{\dagger}$) is annihilation (creation) operator for the photons absorbed (or emitted) due to atomic collisions at the equilibrium, $N$ is a number of atoms. 

The thermodynamic property of a coupled atom-light system  depends   on the contribution of photonic and atomic parts in Eq.\eqref{eq7_} which is usually  considered  in some specific limits. 

The so-called low density limit $\rho \ll 0.5$ for atom-field excitations (polaritons) implies that  
\begin{subequations} \label{eq8_}
\begin{equation*}
\lambda _{}^{2}\ll 1, \ \ \sigma _{bb} \ll \sigma_{aa}\simeq 1,  \eqno{(\ref{eq8_}\rm{a,b})}
\end{equation*}
\end{subequations}
which is obtained at negative atom-field detuning $\delta < 0$ at ``low temperature'' limit \eqref{eq3_}, when atoms mostly populate their ground state $\left| a \right\rangle $.     

Relation (\ref{eq8_}a) represents the necessary prerequisite for observing a second order phase transition in various physical systems being at  low temperatures, cf.~\cite{7,25,27}. As for the problem of OCs a low density limit \eqref{eq8_} can be achieved at high temperatures but for a very large atom-field detuning $\left|\delta \right|$. An experimentally accessible relative photon number involved in the atom-field interaction  and estimated for Rabi frequency $\Omega _{0} /2\pi =0.1$~THz  is  ${N_{ph} / N} \simeq 6.24\times 10^{-3} $ -- see~\cite{14,15}. We assume that inequality (\ref{eq8_}a) is fulfilled for the system under discussion.

The saturation of atomic population is achieved at excitation  density $\rho \approx 0.5$  for $\hbar \left|\delta \right|\ll k_{B} T$. This is the so-called  secular approximation (see~\cite{8}) for which OCs tend  to equalize dressed-state populations which is not the case of our study here.

For positive detuning ($\delta > 0$) from \eqref{eq7_} and (\ref{eq8_}a) we obtain $\rho > 0.5$ ($\sigma_{bb} > 0.5$) that corresponds to the limit of population inversion in a two-level atomic ensemble, cf.~\cite{12}. Physically, such a system behaves  unstable because of spontaneous emission  processes  from the upper level. In particular, relevant  population $\sigma _{bb}$ diminishes at a large positive detuning for any fixed laser intensity, see Fig.\ref{fig1}.

To find photonic field properties under OC process at equilibrium we consider \eqref{eq7_} as an equation for  order parameter  $\lambda $ at the given density $\rho $. Putting in \eqref{eq7_} $\lambda =0$ for critical  value $\alpha_{c}$ of  vital parameter $\alpha \equiv {\hbar \delta / k_{B} T}$ that determines the phase boundary    between  normal  and ``superradiant'' states  one can obtain:
\begin{equation} \label{eq9_} 
\alpha _{C} =-{\rm ln}\left[ \left( {1-\rho }\right) /  {\rho } \right].                                                      
\end{equation} 
Furthermore,  one can get from \eqref{eq6_} -- \eqref{eq8_} an expression for order parameter $\lambda (\alpha )$ as a function of  atom-field detuning $\delta $ or temperature  $T$ 
\begin{equation} \label{eq10_} 
\lambda (\alpha )=\lambda _{\infty } \left[1-\frac{1}{\rho \left[1+\left({1\mathord{\left/ {\vphantom {1 \rho }} \right. \kern-\nulldelimiterspace} \rho } -1\right)^{{\alpha \mathord{\left/ {\vphantom {\alpha  \alpha _{C} }} \right. \kern-\nulldelimiterspace} \alpha _{C} } } \right]} \right]^{1/2},                                     
\end{equation} 
where $\lambda _{\infty } \equiv \sqrt{\rho } $ is an order parameter  at ``zero temperature'' limit \eqref{eq3_}. 

Eq.\eqref{eq10_} gives an opportunity to interpret critical properties of the photonic field occurring due to OCs as a result of atom-field thermalization.  To be more specific we suppose atom-light detuning  $\delta$ to be negative.  While  modulus $\left|\delta \right|$ is small enough  (such as  $\hbar \left|\delta \right|\ll k_{B} T,{\rm \; }\gamma$),  the elementary processes of emission (or absorption) of photons by atoms happen independently on the atomic collisions with buffer gas particles during their  free  motion.  The atomic collisions lead to the dephasing of emitted radiation.   In this sense they are worse and we deal with  a ``normal'' (incoherent) state for photons.
\begin{figure}
\includegraphics[scale=0.25]{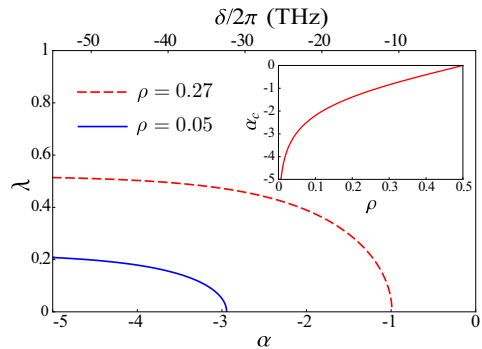}
\caption{\label{fig2} (Color online) Order parameter $\lambda$ versus vital parameter $\alpha$ (normalized atom-light  detuning  $\delta $) for $T=530 $ K temperature of atomic gas. In the inset a phase boundary for critical parameter $\alpha _{C}$ versus LB polariton density $\rho$ is presented (see the text).}
\end{figure}

A physical picture is changed significantly by increasing atom-field detuning $\delta$ at high   buffer gas pressures.  For large  $\left|\delta \right|$ under the condition of $\hbar \left| \delta \right| \gtrsim k_{B}T$  it is no longer possible to ignore correlations between elementary acts of atomic collisions  with buffer gas particles and photon emission (or absorption).  This is a case of OCs. Each collision happens in a  very  short  time period  comparing with the time interval separating two collisions. Watching  OCs for a long time when many collisions happen and  frequent transitions of atomic population between dressed states occur it is possible to create   some population of coupled atom-light states in thermodynamic equilibrium  due to the  thermalization process.  This population can be established for ``usual'' (bare) atomic levels  only for large values of atom-field detuning.  In this case  Eq.\eqref{eq8_}  represents transition to  some ordering (``superradiant'') state  as a result of forming macroscopic  ``spontaneous'' polarization for a coupled atom-light system. It should be noticed, that the existence of some small residual polarization of the atomic system has been predicted in \cite{15} as a result  of the atom-light (dressed) states thermalization.

Figure \ref{fig2} demonstrates phase transition behavior for order parameter $\lambda $ for the system under discussion.   From practical point of view it is much easier  to vary atom-field detuning  $\delta$ instead of atomic gas temperature $T$. The dashed curve corresponds to density  ${\rho={\rm 0.27}}$ in accordance with experimentally achieved detuning $\delta /2\pi =-11 $ THz. The solid curve is relevant to low density limit \eqref{eq8_}. The critical value $\left|\alpha _{C} \right|$ of parameter $\alpha $ is increased by  decreasing excitation density $\rho $ -- see inset in Fig.\ref{fig2}. In the low density limit \eqref{eq8_} Eq.\eqref{eq10_} is simplified and looks like $\lambda (\alpha )\simeq \lambda _{\infty } \left[1-\left(\rho \right)^{{\alpha /\alpha _{C} } -1} \right]^{1/2}$. Phase transition occurs for large enough detuning (or at lower temperatures). 

For positive detuning ($\delta > 0$) the problem of phase transition under discussion is sophisticated. Formally, full thermalization of coupled atom-light states and relevant phase transition can be achieved at infinite resonant Rabi splitting frequency $\Omega_{0}$ and atom-field detuning $\delta$, when the population of the upper level is $\sigma_{bb}=1$ -- see \eqref{eq6_}. However, an approach to OCs that characterizes thermalization process is valid only for limited values of detuning such as $\left| \delta \right| \ll \omega_{at}$. For any finite $\Omega_{0}$ the variation of atom-field detuning $\delta$ or gas temperature $T$ drives a coupled atom-light  system out of thermal equilibrium  even if such an equilibrium (or quasi-equilibrium)  has been initially achieved -- Fig.\ref{fig1}.  In this sense, the fulfillment of condition \eqref{eq5_} for coupled atom-light states permits to create a population inversion in a two-level atomic ensemble and to realize non-equilibrium (or quasi-equilibrium) transition to lasing as a result, cf.~\cite{12, 13}.

\section{PHASE TRANSITION IN A COUPLED ATOM-LIGHT SYSTEM UNDER EQUILIBRIUM}\label{sec_III}

While OCs result in thermalization of coupled (dressed) atom-light states, the Eq.\eqref{eq10_} describes the phase transition  under discussion only  \textit{qualitatively}. Strictly speaking, a rigorous thermodynamical approach to critical properties of a coupled atom-light system at equilibrium is needed. 

We describe the interaction of a single-mode light field $f$ with an atomic ensemble  by relevant (Dicke) Hamiltonian -- cf.~\cite{15,16}.
\begin{equation} \label{eq11_} 
H=\hbar \omega _{L} f_{}^{\dag } f_{} +\frac{\hbar \omega _{at} }{2} \sum _{j=1}^{N}S_{z,j}  +\frac{\hbar \kappa }{\sqrt{N} } \sum _{j=1}^{N}\left(S_{-,j}^{\dag } f+f_{}^{\dag } S_{-,j} \right),                               
\end{equation} 
where $\kappa =g\sqrt{N} $ is a collective parameter of atom-field interaction, $S_{-,j} $ is a transition operator for $j$-th atom, $S_{z,j} $ is an operator of atomic population imbalance.

From the mathematical point of view for the definition of $S_{-,j}$ and $S_{z,j}$ operators it is possible to use Pauli spin matrices representation (cf.~\cite{17}) or explore annihilation ($a_j$, $b_j$) and creation ($a^{\dagger}_j$, $b^{\dagger}_j$) operators for bosonic atoms at the ground $\left| a \right\rangle$ and excited $\left| b \right\rangle$ states  in the second quantization representation respectively. The latter one enables definitions $S_{-,j}=a^{\dagger}_j b_j$ and $S_{z,j} = b^{\dagger}_j b_j - a^{\dagger}_j a_j $ -- see~\cite{28}.

Since the atomic medium is very dense ($n_{at} \simeq 10^{16} cm^{-3}$) one can neglect inhomogeneous (Doppler) broadening  because Doppler broadening doesn't restrict the thermalization process and is by two orders smaller  than  parameter $\kappa$, that is $\kappa /2\pi \approx 0.624$ THz for resonant Rabi frequency $\Omega_{0}/2\pi = 0.1$ THz -- cf.~\cite{14,15}. Hence, the so-called strong atom-field coupling condition that plays an important role in the problem of phase transition with polaritons is fulfilled in our case, cf.~\cite{3,4,5}. In this limit one can introduce collective atomic ladder operator $S_{-}$ and operator of atomic population imbalance $S_z$ as follows:
\begin{subequations} \label{eq12_}
\begin{equation*}
S_{-}=\sum\limits^{N}_{j=1}S_{-,j}, \ \ 
S_{z}=\sum\limits^{N}_{j=1}S_{z,j}. \eqno{(\ref{eq12_}\rm{a,b})}
\end{equation*}
\end{subequations}

Operators introduced in \eqref{eq12_} obey SU(2) algebra commutation relations
\begin{subequations} \label{eq12}
\begin{equation*}
\left[S_{-}^{} ,S_{-}^{\dag } \right]=-S_{z}, \ \ 
\left[S_{z}^{} ,S_{-}^{} \right]=-2S_{-}. \eqno{(\ref{eq12}\rm{a,b})}
\end{equation*}
\end{subequations}

For describing excitations of a two-level atomic system we define annihilation ($\phi$) and creation ($\phi^{\dagger}$) excitation operators by using the so-called Holstein-Primakoff transformation (cf.~\cite{28_})
\begin{subequations} \label{HP}
\begin{equation*}
S_{-}=\sqrt{N-\phi^{\dagger} \phi} \phi, \ \ 
S_{-}^{\dagger}=\phi^{\dagger} \sqrt{N-\phi^{\dagger} \phi}, \ \ 
S_{z}=2\phi^{\dagger}\phi -N. \eqno{(\ref{HP}\rm{a,b,c})}
\end{equation*}
\end{subequations}

The relations \eqref{eq12} are preserved when operators $\phi$ and $\phi^{\dagger}$ obey usual commutation relation for the Bose-system:
\begin{equation}\label{commutation}
\left[ \phi, \phi^{\dagger} \right]=1.
\end{equation}

For a large number of atoms $N$ it is possible to treat operators $S_{-}$ and  $S_{-}^{\dagger}$ as:
\begin{subequations} \label{eq16}
\begin{equation*}
S_{-}=\sqrt{N}\phi - \frac{\phi^{\dagger}\phi^{2}}{2\sqrt{N}}, \ \ 
S_{-}^{\dagger}=\sqrt{N}\phi^{\dagger} - \frac{{\phi^{\dagger}}^{2}\phi}{2\sqrt{N}}. \eqno{(\ref{eq16}\rm{a,b})}
\end{equation*}
\end{subequations}

The last term in \eqref{eq16} characterizes nonlinear effects for atomic excitations under the atom-field interaction. However, in the low density limit defined as
\begin{equation}\label{eq17}
 \phi^{\dagger} \phi \ll N
\end{equation}
we can  obtain from \eqref{eq16} simple expressions for operators $\phi$ and $\phi^{\dagger}$, i.e. we get
\begin{subequations} \label{eq18}
\begin{equation*}
P \equiv \phi = \frac{1}{\sqrt{N}}S_{-} , \ \ 
P^{\dagger} \equiv \phi^{\dagger} = \frac{1}{\sqrt{N}}S_{-}^{\dagger}. \eqno{(\ref{eq18}\rm{a,b})}
\end{equation*}
\end{subequations}

Taking into account Eq.(\ref{HP}c), it is easy to see that condition \eqref{eq17} exactly corresponds to the relation (\ref{eq8_}b) obtained for large and negative atom-field detuning $\delta < 0$. In this case collective atomic excitations described by operator $\phi$ characterize a macroscopic polarization $P$ of atomic system.

Taking into account Eqs.(\ref{HP}c), \eqref{eq17}, Hamiltonian \eqref{eq11_} can be easily reduced to the form
\begin{equation}\label{eq19}
H=\hbar \omega _{L} f_{}^{\dag } f_{} +\hbar \omega _{at}\phi^{\dagger}\phi   +\hbar \kappa\left(  \phi^{\dagger} f+f^{\dagger} \phi \right).
\end{equation}

At the steady-state for a coupled atom-light system Hamiltonian \eqref{eq19} can be diagonalized with the help of unitary transformations
\begin{subequations} \label{eq13_}
\begin{equation*}
\Phi _{1} =\vartheta _{1} f+\vartheta _{2} \phi, \ \ 
\Phi _{2} =\vartheta _{1} \phi  -\vartheta _{2} f, \eqno{(\ref{eq13_}\rm{a,b})}
\end{equation*}
\end{subequations}
where  $\vartheta _{1,2}^{2} =\frac{1}{2} \left(1\pm \frac{\delta }{\sqrt{\delta ^{2} +4\kappa ^{2} } } \right)$ are Hopfield coefficients that obey normalization condition $\vartheta _{1}^{2} +\vartheta _{2}^{2} =1$.

The annihilation operators $\Phi _{1,2}^{} $   in Eq.(\ref{eq13_}a,b) characterize two types of quasiparticles due to the atom-field interaction, i.e. upper and lower branch polaritons. By using relation \eqref{commutation} it is easy to show that operators defined in (\ref{eq13_}a,b) obey commutation relations
\begin{equation}\label{eq14_}
\left[\Phi _{1}^{} ,\Phi _{1}^{\dag } \right]=\left[\Phi _{2}^{} ,\Phi _{2}^{\dag } \right]=1.
\end{equation}

Below we focus on photon-like polaritons necessary for explaning photonic phase transition for the light field in the system under discussion. 

The equation for order parameter \textit{ }$\lambda $\textit{ }at equilibrium can be derived from a variational (thermodynamic) approach, see, for example, \cite{17}. In their calculations the authors use a canonical ensemble with chemical potential  $\mu =0$ applied to Dicke Hamiltonian \eqref{eq11_}. However, this approach cannot be   justified even for pure photonic system posing phase transition at equilibrium. For example, as it is shown in \cite{29} the chemical potential for photonic gas being under BEC condition is nonzero. 

In the paper we  use a grand canonical ensemble with finite  chemical potential and  coherent basis \textit{ }for the photonic field for calculating partition function $Z(N,T)=Tr\left(e^{-{H'\mathord{\left/ {\vphantom {H' k_{B} T}} \right. \kern-\nulldelimiterspace} k_{B} T} } \right)$; $H'=H-\mu N_{ex} $ is a modified Hamiltonian; $N_{ex} =f_{}^{\dag } f_{} +\frac{1}{2} \sum \limits _{j=1}^{N}S_{z,j}  $ is  the number of excitations. 

Evaluating partition function $Z(N,T)$ under the  mean field approximation one can obtain (cf.~\cite{17}):
\begin{equation} \label{eq16_} 
\tilde{\omega }_{L} \lambda =\frac{\kappa ^{2} \lambda \tanh \left(\frac{\hbar }{2k_{B} T} \left(\tilde{\omega }_{at}^{2} +4\kappa _{}^{2} \lambda _{}^{2} \right)^{1/2} \right)}{\left(\tilde{\omega }_{at}^{2} +4\kappa _{}^{2} \lambda _{}^{2} \right)^{1/2} }  ,        
\end{equation} 
where we introduced denotations $\tilde{\omega }_{L} \equiv \omega _{L} -\mu $ , $\tilde{\omega }_{at} \equiv \omega _{at} -\mu $.

Equation \eqref{eq16_} is similar to the gap equation that characterizes a second-order phase transition for various systems in solid-state physics, see e.g.~\cite{27}. For finding chemical potential $\mu $ in our case one can use polariton number operator $N_{pol}=\Phi^{\dagger}_{1}\Phi_{1}+\Phi^{\dagger}_{2}\Phi_{2}$. With the help of Eqs.(\ref{eq12_}b), (\ref{HP}c) and \eqref{eq13_} $N_{pol}$ can be represented as
\begin{equation}\label{N}
N_{pol}=f_{}^{\dag } f+\sum\limits_{j=1}^{N} b_{j}^{\dagger}b_{j}= \frac{N}{2}+N_{ex}. 
\end{equation}
Density $\rho $ of  total excitations occurring in a closed atom-light field  system   defined in \eqref{eq7_} can be interpreted as  a  polariton number density, i.e. $\rho \equiv N_{pol} / N$. In the thermodynamic limit it approaches 
\begin{equation} \label{eq17_} 
\rho =\lambda _{}^{2} +\frac{1}{2} \left[1-\frac{\tilde{\omega }_{at}^{} \tanh \left(\frac{\hbar }{2k_{B} T} \left(\tilde{\omega }_{at}^{2} +4\kappa _{}^{2} \lambda _{}^{2} \right)^{1/2} \right)}{\left(\tilde{\omega }_{at}^{2} +4\kappa _{}^{2} \lambda _{}^{2} \right)^{1/2} } \right].       
\end{equation} 

The last term in the brackets of Eq.\eqref{eq17_} characterizes normalized population imbalance $\tilde{S}_z = S_z / N$ at thermal equilibrium. In particular, it is (cf.~\cite{25}):
\begin{equation}\label{S_z_eq}
\tilde{S}^{(eq)}_z = \frac{\tilde{\omega }_{at}^{} \tanh \left(\frac{\hbar }{2k_{B} T} \left(\tilde{\omega }_{at}^{2} +4\kappa _{}^{2} \lambda _{}^{2} \right)^{1/2} \right)}{\left(\tilde{\omega }_{at}^{2} +4\kappa _{}^{2} \lambda _{}^{2} \right)^{1/2} }.
\end{equation}

Combining \eqref{eq17_} and \eqref{eq16_} for chemical potential $\mu $ we get
\begin{equation} \label{eq18_} 
\mu _{1,2} =\frac{1}{2} \left[\omega _{at} +\omega _{L} \pm \Omega _{R,eff} \right],                                                   
\end{equation} 
where $\Omega _{R,eff} =\sqrt{\delta _{}^{2} -8\kappa ^{2} \left(\rho -\lambda _{}^{2} -\frac{1}{2} \right)} $ is a new effective Rabi splitting frequency. 

At low polariton  densities  (see \eqref{eq8_}) Eq. \eqref{eq18_}  defines normal state ($\lambda =0$) for  upper ($\mu _{1}$)  and lower ($\mu _{2} $)  polariton branch frequencies respectively.  Inserting \eqref{eq18_} into \eqref{eq16_} we arrive at 
\begin{equation} \label{eq19_} 
\tilde{\omega }_{L1,2}^{} =\frac{\kappa ^{2} \tanh \left(\frac{\hbar }{2k_{B} T} \left(\tilde{\omega }_{at1,2}^{2} +4\kappa _{}^{2} \lambda _{}^{2} \right)^{1/2} \right)}{\left(\tilde{\omega }_{at1,2}^{2} +4\kappa _{}^{2} \lambda _{}^{2} \right)^{1/2} },
\end{equation} 
that are BCS-like equations  for upper (index ``1'') and lower (index ``2'') branch polaritons respectively; $\tilde{\omega }_{L1,2}^{} =\frac{1}{2} \left[\delta \mp \Omega _{R,eff} \right]$, $\tilde{\omega }_{at1,2}^{} =\frac{1}{2} \left[-\delta \mp \Omega _{R,eff} \right]$.                          

The critical temperature $T_{C}$ of phase transition (for given atom-field detuning $\delta$) can be obtained from \eqref{eq19_} for $\lambda =0$ and looks like  
\begin{equation} \label{eq20_} 
T_{C1,2} =\frac{\hbar \left|\tilde{\omega }_{at1,2}^{} \right|}{2k_{B} {\rm \; tanh}^{-1}\left[\pm \left( 2\rho -1\right) \right]}.
\end{equation}

The results, obtained from Eq. \eqref{eq20_} are consistent with  existing  theories of BCS-like phase transition in polariton system, cf.~\cite{7,25}. 

For very large atom-light detuning, i.e. for $ \left|\delta \right| \gg \kappa $, a normalized population imbalance \eqref{S_z_eq} looks like 
\begin{equation}\label{eq29_}
\tilde{S}_{z}^{(eq)} \approx -\frac{\delta}{\left| \delta \right|} {\rm tanh} \left( \hbar \left| \delta \right| / 2k_{B}T \right).
\end{equation}
Equation \eqref{eq29_} corresponds to equilibrium population of the upper level achieved at full thermalization of coupled atom-light states in the presence of OCs and described by Eq.\eqref{eq6_}. In the same limit the critical temperature $T_{c}$ approaches
\begin{equation}\label{eq30_}
T_c = \frac{\hbar \delta}{2 k_{B} {\rm tanh}^{-1}(2\rho -1)}.
\end{equation}
Eq.\eqref{eq30_} immediately comes from Eq.\eqref{eq7_} taken for order parameter $\lambda = 0$.

At the same time Eq.\eqref{eq20_} defines  critical value $\alpha _{C} $ of  $\alpha $-parameter (normalized atom-field detuning  $\delta$)  for fixed temperature $T$ of atomic gas   for which we get  
\begin{equation} \label{eq21_} 
\alpha _{C} =-\ln \left(\frac{1-\rho }{\rho } \right)-\frac{2\hbar ^{2} \kappa ^{2} \left(\rho -1/2\right)}{k_{B}^{2} T^{2} \ln \left({(1-\rho )\mathord{\left/ {\vphantom {(1-\rho ) \rho }} \right. \kern-\nulldelimiterspace} \rho } \right)}.  
\end{equation} 

Equation \eqref{eq21_} characterizes  phase boundary  between photonic superradiant (coherent) and normal states  taken at $\lambda =0$ for the given temperature $T$ of an atomic ensemble. In accordance with experimental conditions of \cite{14,15,32},  $k_{B} T \gg \hbar \kappa$; and the last term in \eqref{eq21_} vanishes. In this case we recognize Eq.\eqref{eq9_}  for critical  parameter $\alpha _{C} $  that characterizes OC process at full equilibrium.  

For negative atom-light detuning, i.e. for $\delta < 0$, we can interpret the phase transition under consideration as a transition to superfluid state of LB photon-like polaritons ($\Phi _{2} \simeq -f$) occurring under OC process in the low density limit \eqref{eq8_}. In this case order parameter $\lambda$ can be recognized as a polariton wave function. In Appendix we represent the main features of cylindrical waveguides for observing phase transition with polaritons. Superfluid properties of polaritons can be elucidated taking into account a weak interaction between them and can be described in the momentum representation, cf.~\cite{4}. Such an interaction arises due to nonlinear effects in atom-field interaction, described by second terms in expressions \eqref{eq16} for $S_{-}$ and $S_{-}^{\dagger}$, cf.~\cite{30}.  We expect, that in the future it will be possible  to examine superfluid properties of such polaritons by using very thin-walled (up to 60 nm) waveguides for the experiment with Josephson junctions at high enough temperatures.

\section{CONCLUSIONS AND OUTLOOK}

We consider a phase  transition problem   in a two-level atomic ensemble strongly interacting with an optical field.  The main features of such a phase transition are connected with a coupled atom-light state thermalization occurring due to the OCs process with buffer gas particles. Such a thermalization can be achieved experimentally for a large value of negative atom-field detuning $\delta$. Using a thermodynamic approach (partition function) we established a gap equation under a mean field approximation for order parameter $\lambda$ -- normalized average optical field amplitude. In the paper we represent simple arguments confirming that the obtained thermalization of coupled atom-light states leads to the photonic phase transition to the superradiant (coherent)  state and is characterized by some ordering (equilibrium state)  for two-level atomic system. Nontrivial solution of Eq.\eqref{eq16_} with $\lambda \neq 0$ leads to the appearance of a macroscopic stationary polarization of the atomic medium that is proportional to the order parameter $\lambda$. 

The equilibrium  properties of a coupled atom-light system offer a promising approach to studying the critical behavior of photon-like LB polaritons under the low density limit \eqref{eq8_}.  In this respect, we discuss special metallic waveguides for providing necessary atom-field interaction and achieving suitable polariton lifetime, as a result. In particular,  phase transition under discussion can be connected with  transition to superfluid (coherent) state for such polaritons characterizing superposition of a quantized optical field and a macroscopic atomic polarization.   In the presence of phase transition the atomic polarization evolves in time with the frequency that is equal to the chemical potential $\mu_2 \approx \omega_{at}- \left| \delta \right|$ for LB polaritons. This property enables to observe the phase transition under discussion experimentally.

We note that the results obtained above can be useful for observing true BEC occurring with atomic polaritons. In this case we need to realize certain trapping potential for LB polaritons. One simple way for that is to use a biconical waveguide cavity for which waveguide radius $R$ varies smoothly along $z$-coordinate. A similar dielectric (bottle like) cavity based on tapered optical fiber has been recently proposed for improving atom-light coupling strength, see e.g.~\cite{31}. In our case such a cavity enables to create an appropriate  trapping potential for photons as well as for polaritons in $z$-dimension. It is important that in this case the lifetime of photon-like polaritons can be long enough and is determined by the cavity Q-factor. These problems will be a subject of intensive study both in theory and experiment in forthcoming papers.

This work was supported by RFBR Grant No. 09-02-91350 and by Russian Ministry of Education and Science under Contracts No. $\Pi$569, 14.740.11.0700. A. P. Alodjants is grateful to M. Weitz and F. Vewinger for hospitality and valuable discussions. We are grateful to the referee for his comments.

\appendix

\section{PHOTON CONFINEMENT IN CYLINDRICAL WAVEGUIDE}

Let us briefly discuss the possibility to observe the phase transition under discussion with LB polaritons in metallic cylindrical waveguide. The implementation of metallic microtubes for investigating  coupled atom-light state thermalization under OC has been demonstrated in \cite{32}. The properties of an optical field in the  ideal empty waveguide without any losses are well known -- see e.g.~\cite{33}.  In particular,  wave vector  component $k_{\bot ,mp} ={g_{mp} / R} $ of the field  in the cross section of the waveguide is orthogonal to $z$-axis and quantized; $g_{mp} $ represents the root of the equation $J_{m} (k_{\bot } R)=0$, $R$ is a waveguide radius,  $m$  and $p$ are integer numbers characterizing azimuthal  and transversal field distribution respectively. At the same time there exist  continuum modes with wave vectors $k_{z} $ in $z$ direction. Physically it means that it is  possible to establish a dispersion relation for photonic field in the waveguide  as
\begin{equation} \label{eqA1_} 
\omega _{L} \simeq ck_{\bot ,mp}^{} +{\hbar k_{z}^{2}}/{2m_{ph} },                           
\end{equation}                                                          
where we have  introduced photon mass $m_{ph} ={\hbar k_{\bot ,mp}/c}$. Thus, for a cylindrical waveguide spatial degrees of freedom in $x$ and $y$ directions are suppressed and photons remain confined in the plane which is perpendicular  to $z$-axis. At the same time Eq.\eqref{eqA1_} implies the existence of a finite polariton mass in the waveguide. For example, mass $m_{pol}$ of photon-like polaritons is about  $2.6\times 10^{-36}$~kg that implies a high temperature phase transition in our case, cf.~\cite{3,4,5}.

It is possible to define from Eq.\eqref{eqA1_} the so-called \textit{cutoff wave number} $k_{mp}^{(c)} = k_{\bot , mp} = {g_{mp}} / R $ for which  a waveguide mode propagation constant vanishes.  In the experiment it is preferable to use a fundamental $\mathrm{TM_{01}}$-mode with quantum numbers $m=0$, $p=1$ ($g_{01} =2.4048$) as a cavity mode for the atom-field interaction. In particular, we require the fulfillment of conditions  $\mu _{1,2} ,\omega _{L} >ck_{01}^{(c)} $ for  characteristic frequencies that describe a coupled atom-light system.  The condition under discussion can be represented in some other way by using   radius $R$ of the waveguide  as 
$c\frac{g_{01} }{\omega _{at} -|\delta |} <R<c\frac{g_{11} }{\omega _{at} +|\delta |}$, where $g_{11}\simeq 3.8317$. The fulfillment of the last inequality guarantees that  only a single waveguide  mode effectively interacts with the atomic ensemble at a large atom-light detuning  $\delta$. Notably, that according to the condition presented above the diameter of the waveguide should be approximately of the order of  wavelength $\lambda _{Rb} \simeq 785$ nm  for rubidium atom \textit{D}-line  transition, cf.~\cite{32}.

\end{document}